\documentclass[10pt]{iopart}
\usepackage{graphicx}
\usepackage{dcolumn}
\usepackage{bm}
\usepackage{color}
\usepackage{xcolor}
\usepackage{tikz}
\usepackage{amssymb}

\newcommand{\new}[1]{\textcolor{black}{#1}}
\newcommand{\refone}[1]{\textcolor{black}{#1}}
\newcommand{\reftwo}[1]{\textcolor{black}{#1}}
\newcommand{\remark}[1]{\textcolor{black}{#1}}

\begin{document}

\title[Reduced turbulent transport in the stellarator configuration CIEMAT-QI4]{Reduced turbulent transport in the quasi-isodynamic stellarator configuration CIEMAT-QI4}

\author{J.~M.~García-Regaña, I.~Calvo, E.~Sánchez, H.~Thienpondt, J.~L.~Velasco and J.~A.~Capitán}
\address{Laboratorio Nacional de Fusión, CIEMAT, 28040 Madrid, Spain}
\ead{jose.regana@ciemat.es}
\vspace{10pt}
\begin{indented}
\item[]August 2017
\end{indented}

\begin{abstract}
	CIEMAT-QI4 is a quasi-isodynamic stellarator configuration that simultaneously features very good fast-ion confinement in a broad range of $\beta$ values, low neoclassical transport and bootstrap current, and ideal magnetohydrodynamic stability up to $\beta=5\%$. In this paper
	it is shown that CIEMAT-QI4 also exhibits reduced turbulent transport. This is demonstrated through nonlinear electrostatic  simulations with the gyrokinetic code \texttt{stella}, including kinetic ions and electrons. 
	The relation between reduced turbulent transport and the fact that CIEMAT-QI4 very approximately satisfies the so-called maximum-$J$ property is discussed.
\end{abstract}

\ioptwocol
\section{Introduction}

\begin{figure*}
	\begin{center}
		\includegraphics{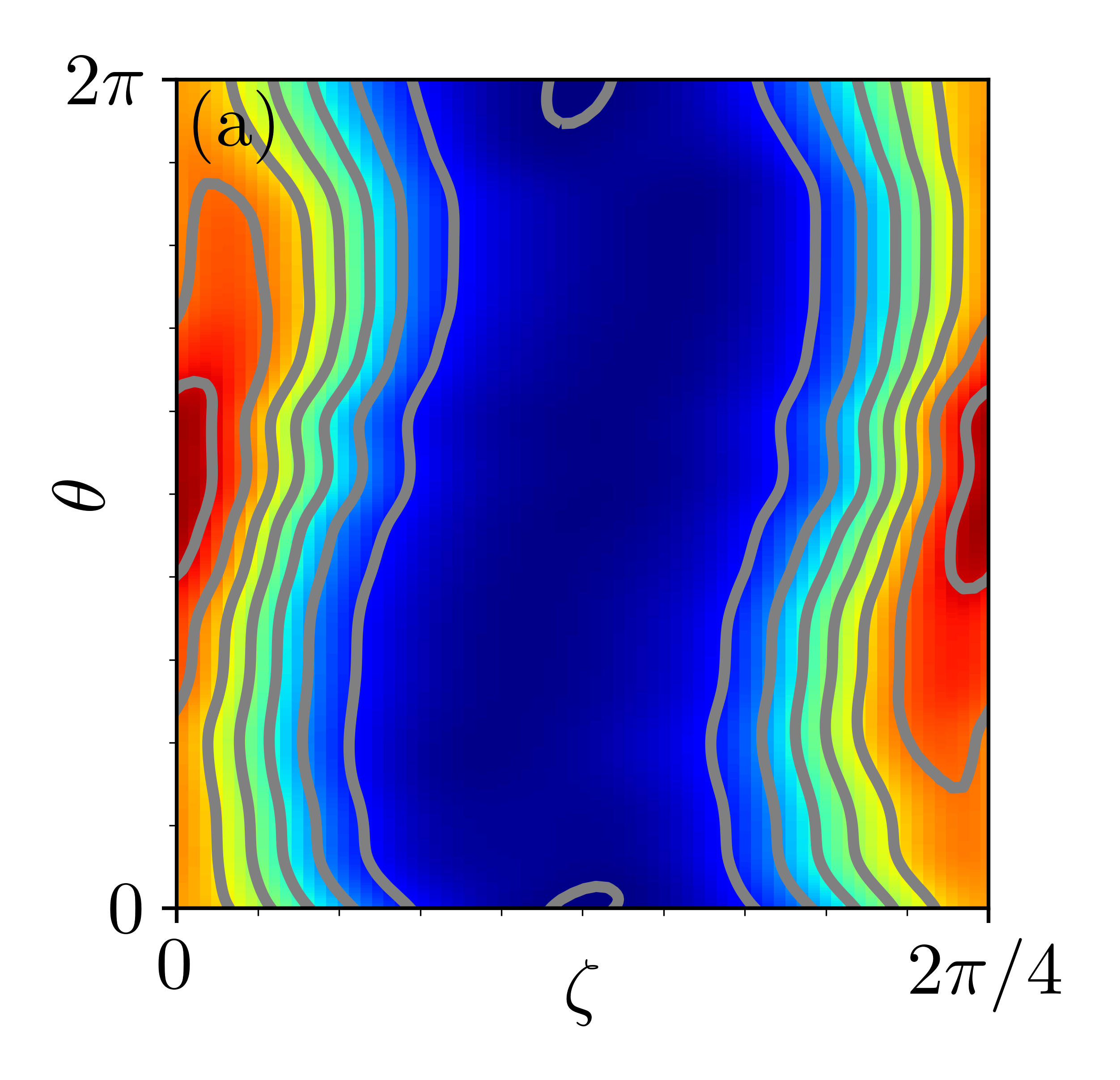}
		\includegraphics{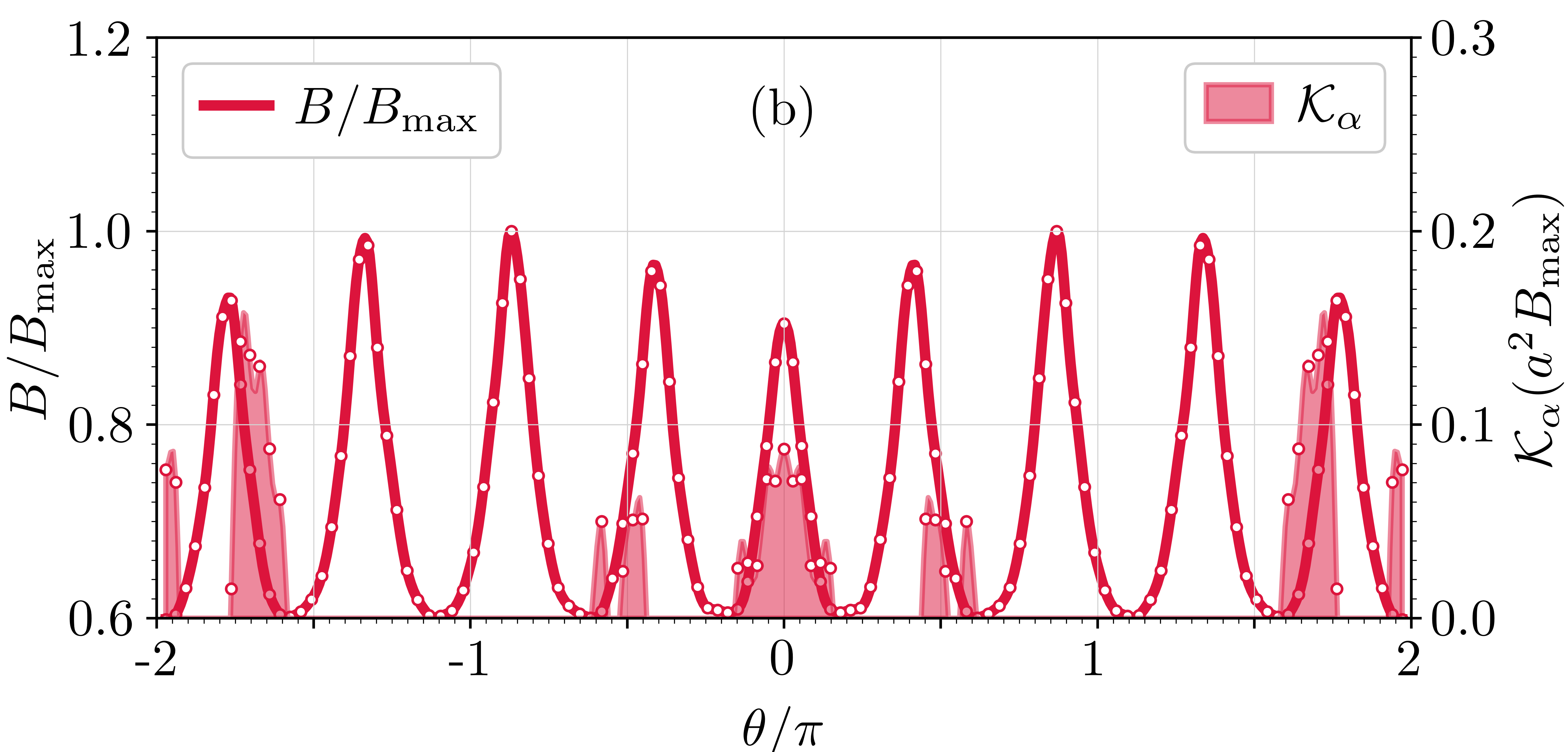}\\
		\includegraphics{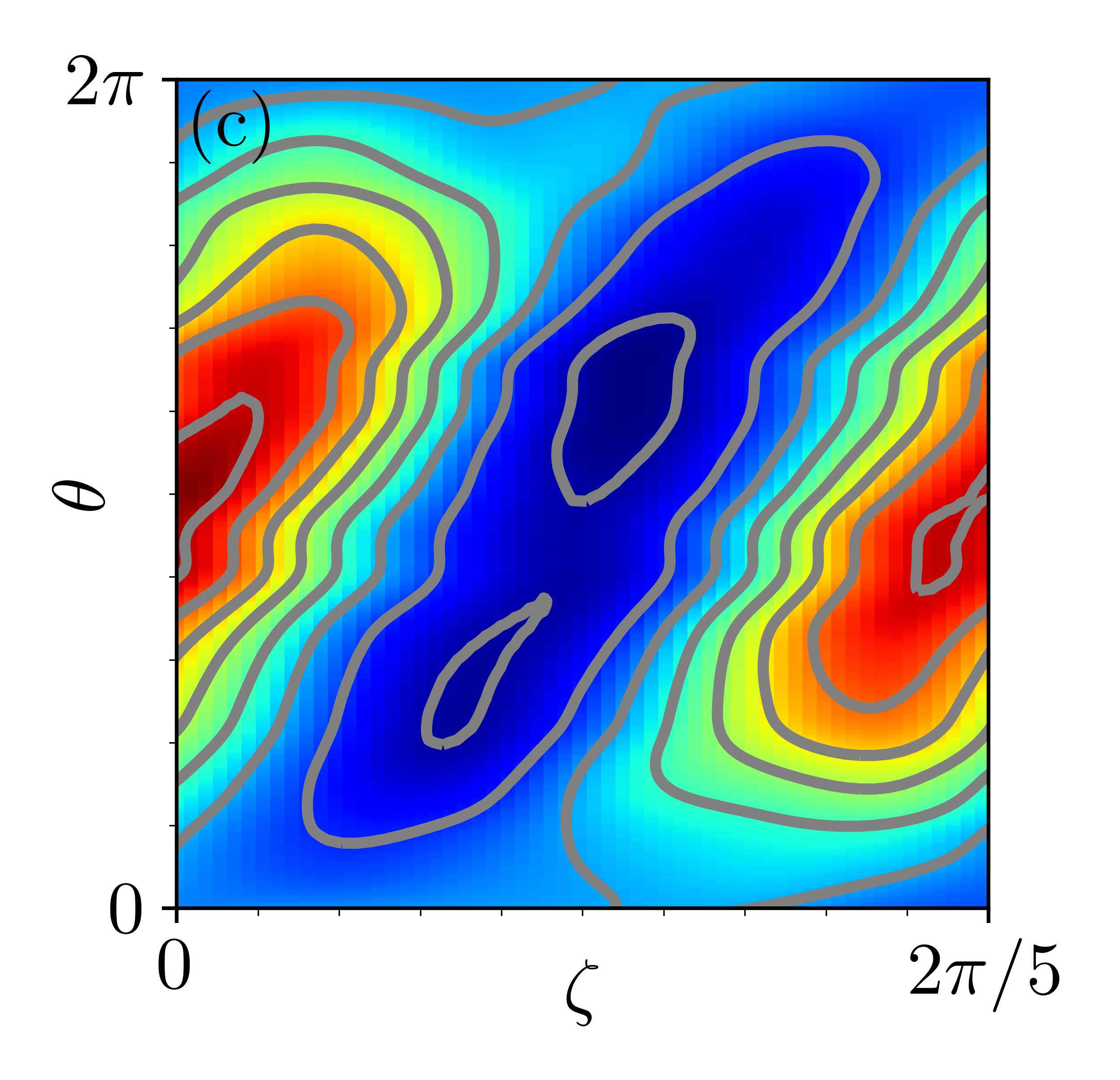}
		\includegraphics{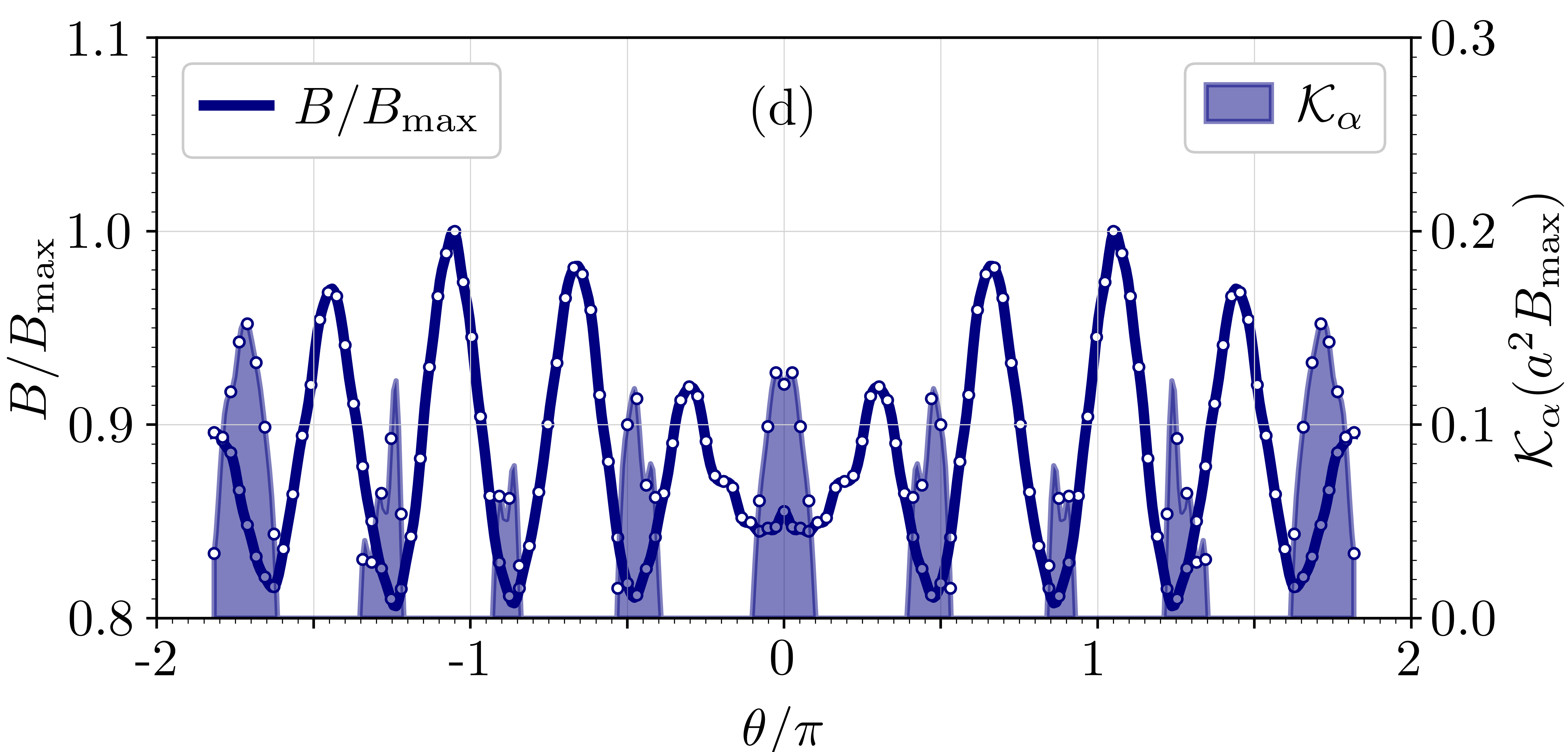}
	\end{center}
	\vspace{-4mm}
	\caption{Magnetic field strength, $B$, of CIEMAT-QI4 (a) and W7-X (c), both at $\beta=1.5\%$ and at $r/a=0.7$,  where red and blue tones represent, respectively, higher and lower values. Magnetic field strength (solid line) and bad curvature r	egions (shaded areas) as a function of the poloidal coordinate for CIEMAT-QI4 (b) and W7-X (d)  \remark{(the markers represent the parallel grid points used in the nonlinear simulations)}. \reftwo{With respect to other differences between configurations, the aspect ratio $A=R_0/a$, with $R_0$ the major radius of the device, rotational transform and global magnetic shear are $\left\{A, \iota, \hat{s}\right\}=\left\{9.94, 0.877, -0.022\right\}$ in the case of CIEMAT-QI4 and  $\left\{A, \iota, \hat{s}\right\}=\left\{10.73, 0.876, -0.110\right\}$ in the case of W7-X.}}
	\label{fig:configurations}
\end{figure*}

In stellarators \cite{Helander_rpp_77_087001_2014}, the magnetic field is \refone{solely} generated by means of external coils. This prevents current-driven
instabilities and makes steady-state operation easier compared with the tokamak, whose axisymmetric magnetic
field is created, in part, by driving a large toroidal current in the plasma. This basic difference on how the
magnetic field is produced is the reason why the stellarator is an attractive potential alternative to the tokamak
concept for fusion power plants. However, \refone{confining a plasma by means of a magnetic field generated with external magnets and no plasma current} implies
that stellarator configurations are inherently three-dimensional and, in general, do not enjoy the excellent
confinement properties that are intrinsic to tokamaks thanks to axisymmetry. \remark{For a stellarator to exhibit
confinement levels comparable to the tokamak}, its magnetic field must be carefully tailored in a process known
as stellarator optimization.

One approach to optimization consists of \remark{looking for three-dimensional magnetic fields such that the magnetic
field strength, in certain sets of coordinates, possesses a symmetry direction. These magnetic fields are called
quasisymmetric. In exactly quasisymmetric stellarators, like in a tokamak, all collisionless orbits are confined.
The Helically Symmetric eXperiment (HSX) \cite{Anderson_ft_1995}} and the design of the National Compact Stellarator Experiment
(NCSX) \cite{Zarnstorff2001} provide examples of stellarator magnetic fields optimized looking for quasisymmetry. More
recently, magnetic fields extremely close to exact quasisymmetry have been obtained \cite{Landreman2022}.

The approach followed in the design of Wendelstein 7-X (W7-X), the first large optimized stellarator, is
different and is based on the concept of quasi-isodynamicity. A magnetic field is quasi-isodynamic if, without
necessarily exhibiting explicit symmetries, it satisfies two properties: all collisionless orbits are confined and
magnetic-field-strength contours are poloidally closed. \refone{Unlike in tokamaks, where particle and heat diffusivity decreases as collisionality decreases, in a non-optimized stellarator, under the worst conditions, neoclassical diffusivity scales inversely with collisionality. In contrast, e}xact quasi-isodynamicity guarantees a neoclassical
transport level similar to that of tokamaks and a vanishing bootstrap current \cite{Landreman_pop_19_2012}. 

The neoclassical optimization of W7-X has been experimentally demonstrated \cite{Beidler_Nature_2021}, representing an enormous
success and a key milestone for the stellarator research programme. The possibility to control the size of the
toroidal plasma current, important to preserve the island structure on which the divertor of W7-X is based, has
also been proven \cite{Dinklage_Nature_2018}. However, the optimization of W7-X 
\refone{has not considered}
two critical aspects \refone{that would ease}
the path towards stellarator reactors: fast-ion confinement and turbulent transport. In a reactor, even a small
fraction of alpha particles lost before thermalization could severely damage the wall, but W7-X was not
designed to confine fast ions at low normalized plasma pressure, $\beta$, and the situation is expected to improve only moderately at high $\beta$ (see e.g.~\cite{Velasco_NF_2021a}). As
for turbulent transport, this was not a feasible optimization target when W7-X was designed, whereas the first
experimental campaigns have shown that turbulence can dominate transport across the entire plasma radius and
can contribute\refone{, among other mechanisms}, to the clamping of the core ion temperature \cite{Beurskens_nf_2021}.

Over the last few years, thanks to theory developments and increased computational capabilities, significant
progress has been made in the quest for the next generation of quasi-isodynamic configurations. A number of
new techniques and codes have been devised to produce configurations that are close to quasi-isodynamicity
\cite{Sanchez_NF_2023, Plunk_JPP_2019,Jorge_JPP_2022,CamachoMata_JPP_2022,Goodman_JPP_2023}.
Among all these efforts, we focus in the present work on the configuration CIEMAT-QI4, 
presented in \cite{Sanchez_NF_2023}.
CIEMAT-QI4 is optimized with respect to fast-ion confinement in a broad
range of $\beta$ values and, at the same time, is ideal MHD stable, and gives low neoclassical transport and bootstrap
current. All these physics aspects are extensively covered in \cite{Sanchez_NF_2023}.
Here, we further characterize the physics performance of this 
configuration by presenting an initial analysis of its turbulent transport.
It is worth noting that CIEMAT-QI4 belongs to the familiy 
of quasi-isodynamic configurations with flat mirror term\footnote{The mirror term for a quasi-isodynamic configuration is defined in \cite{Velasco_NF_2023} as $\sum_{n>0}B_{0n}$, where $B_{mn}$ are the modes of the
	Fourier expansion of the magnetic field strength expressed in Boozer coordinates, $m$ is the poloidal number and $n$ is
	the toroidal number.}, which naturally tend to satisfy the maximum-$J$
property at low $\beta$ \cite{Velasco_NF_2023}. This property entails that the second adiabatic invariant\footnote{The second adiabatic invariant, $J$, is defined for trapped particles as $J =\oint v_{||} {\rm d} l$, where the integral is taken over the trapped orbit, $v_{\|}$ is the parallel velocity and $l$ is the arc length along magnetic field lines.}, $J$, 
is constant on magnetic surfaces and decreases monotonically with the radius. The maximum-$J$ property has 
been argued to be beneficial for mitigating turbulence stemming from trapped electron modes driven 
by density gradients and other ion-scale gyrokinetic instabilities involving kinetic electrons \cite{Rosenbluth_PF_1968, Helander_PoP_2013, Proll_JPP_2022, Plunk_JPP_2017}. Consistent with these expectations, we confirm in the present work that turbulent transport in CIEMAT-QI4 is reduced with respect to W7-X.

The rest of the paper is organized as follows. In section \ref{sec:configurations}, we describe the configuration CIEMAT-QI4. For comparison, specifics of the standard configuration of Wendelstein 7-X (W7-X) at the same value of $\beta$ are also provided. In section \ref{sec:turbulence} the turbulence results for CIEMAT-QI4 and W7-X, obtained with the gyrokinetic code \texttt{stella}, are analysed. For the case of CIEMAT-QI4, the results are expanded by considering three other $\beta$ values, as the maximum-$J$ property is more closely fulfilled with increasing $\beta$. Finally, conclusions are summarised in section \ref{sec:conclusions}.
\section{The magnetic configuration CIEMAT-QI4}
\label{sec:configurations}

\begin{figure*}
	\begin{center}
	\centering
	\includegraphics{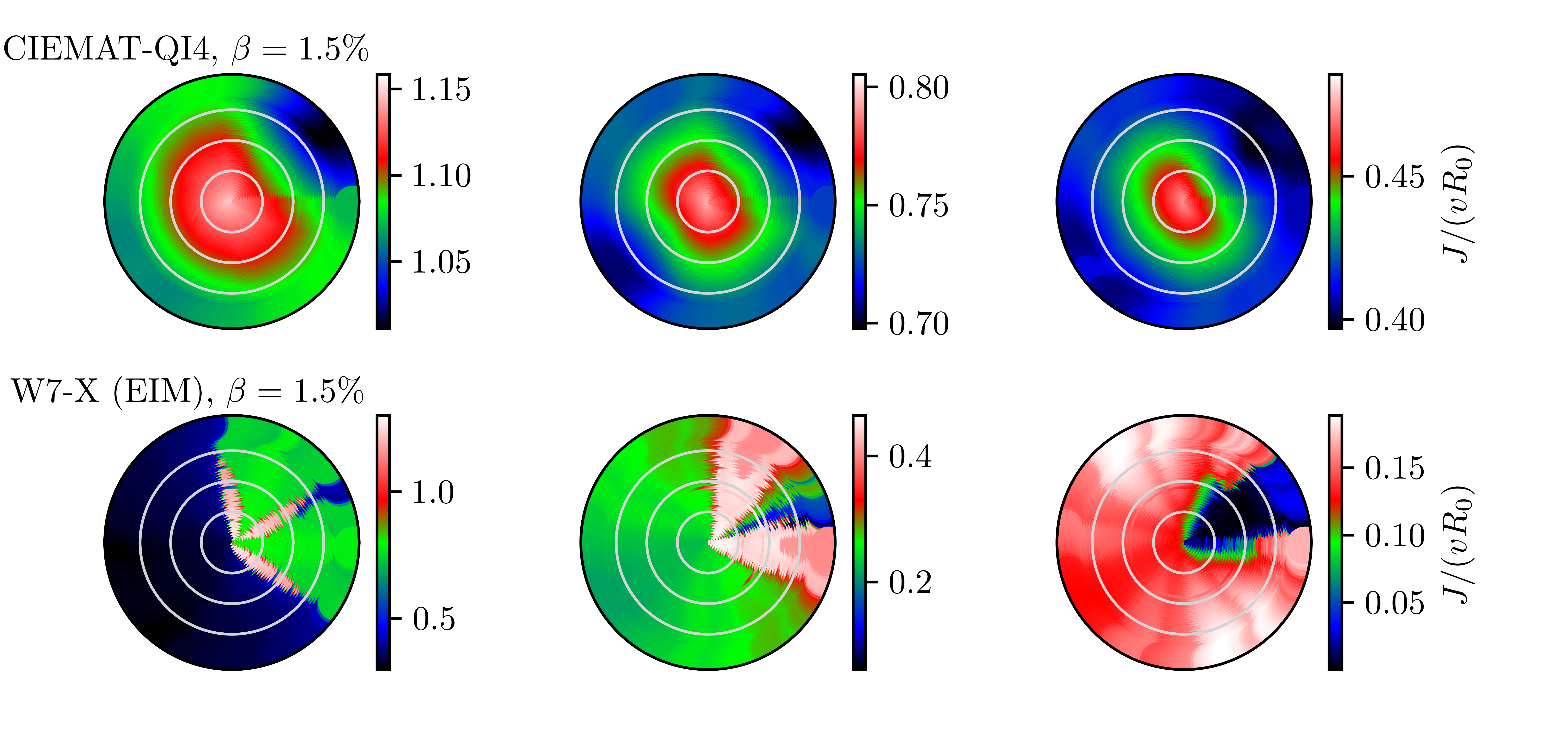}
		\vspace{-4mm}
		\caption{Second adiabatic invariant, $J$, for CIEMAT-QI4 (top row) and W7-X (EIM) (bottom row), both at $\beta=1.5\%$. 
				From left to right, the plots correspond to representative examples of barely (left), moderately (center) and deeply (right) trapped particles \reftwo{with
				ratios of magnetic moment to total energy of $\mu/\mathcal{E}$ equal	to $0.16$, $0.18$ and $0.20$ T$^{-1}$, respectively}.
				For this representation $(r/a)^2$ and $\alpha$ are used as radial and angular, respectively, polar coordinates. The circular
				contours represent the flux surfaces $(r/a)^2=\{0.25, 0.50, 0.75\}$. Note that in this polar representation $\alpha$ is defined in terms of the Boozer, instead of PEST, angular coordinates.}
		\label{fig:maximumJ}
	\end{center}
\end{figure*}

CIEMAT-QI4 is a 4-field-period configuration with aspect ratio $A\approx 10$ that has been obtained using the
optimization suite of codes STELLOPT \cite{Lazerson_stellopt_2020}. 
The code KNOSOS \cite{Velasco_jcp_2020} 
has been integrated into STELLOPT and
employed to compute novel orbit-averaged quantities that are used as proxies for the optimization of fast-ion
confinement \cite{Velasco_NF_2021a}. More details of the optimization runs, target function and  proxies can be
found in \cite{Sanchez_NF_2023}. An aspect not considered explicitly in the STELLOPT optimization loop that led to CIEMAT-QI4 is turbulent transport.
That is the central aspect addressed in this work.

While the characteristics of CIEMAT-QI4 are very good in a
broad range of $\beta$ values, $\beta = 1.5\%$ is the value of $\beta$ for which the configuration was optimised 
(here, $\beta$ is a volume-averaged quantity, obtained considering parabolic 
plasma pressure values). 
For this case, the magnetic field strength $B$ in one period is represented
in figure \ref{fig:configurations}(a) for the radial position $r/a=0.7$, which is considered in the turbulence 
simulations presented in section \ref{sec:turbulence}. 
Here, $r$ is the effective minor radius coordinate and $a$ is its value at the last closed flux surface. In figure \ref{fig:configurations}(b) the 
magnetic field strength is also depicted along the magnetic field line $\alpha=0$, extended approximately two poloidal turns, 
used for the flux tube simulations discussed in section \ref{sec:turbulence}. 
Here $\alpha=\theta-\iota\zeta$ labels magnetic field lines, $\theta$ and $\zeta$ are, respectively, 
the poloidal and toroidal angular flux PEST \cite{Grimm_jcp_49_1983} 
coordinates, the field line is centered with respect to the
point $(\theta, \zeta)=(0,0)$, corresponding to the outboard mid-plane at the bean-shaped poloidal cross section\new{, 
and $\iota$ is the rotational transform}. 
On the other hand,  $\mathcal{K}_{\alpha}=(\mathbf{b}\times\nabla B)\cdot\nabla\alpha/\psi'_t$ is a geometric 
coefficient related to magnetic and curvature drifts. Microinstabilities are prone to localise where $\mathcal{K}_\alpha > 0$, a condition that defines what are usually known as bad curvature regions. Here,
$\mathbf{b}$ is a unit vector pointing in the direction of the magnetic field
and $\psi'_t$ the radial derivative of the toroidal flux. 
For this equilibrium, generated with the code \texttt{VMEC} \cite{Hirshman_pop_28_1985},
the shaded areas in figure \ref{fig:configurations}(b)	
indicate regions of bad curvature.
For comparison, in figures \ref{fig:configurations}(c) and \ref{fig:configurations}(d), the equivalent two plots are presented for the 
standard (or EIM) W7-X configuration, also at $\beta= 1.5\%$.
Within the broad space of W7-X configurations, the standard configuration is generated by a set of modular coils, all of them carrying the same current. This is similar to the case of CIEMAT-QI4, as described in \cite{Sanchez_NF_2023}, where a preliminary coil set is presented. 
The neoclassical optimization of W7-X (one of its main design objectives) has been experimentally demonstrated in the standard configuration, resulting in record values of the fusion triple product in stellarator plasmas~\cite{Beidler_Nature_2021}. \refone{It is assumed that at the value $\beta = 1.5\%$ considered for the comparison, the electrostatic approximation holds}.
	
To the naked eye, by observing the presence of poloidally closed contours in CIEMAT-QI4 in figure \ref{fig:configurations}(a)
and the clear alignment of maxima and minima of $B$ in figure \ref{fig:configurations}(b), in contrast
with the equivalent W7-X figures \ref{fig:configurations}(c) and \ref{fig:configurations}(d),
one can guess the different degree of quasi-isodynamicity of the two configurations.
As for the coefficient $\mathcal{K}_\alpha$, 
while W7-X exhibits a noticeable overlap of magnetic field wells and bad curvature regions, 
usually identified as the origin of trapped-particle-driven instabilities, in CIEMAT-QI4 
the overlap is much smaller.

The maximum-$J$ property is normally invoked in the context of turbulence studies when it comes to arguing the resilience of the configurations to turbulence driven by trapped-electron-modes (TEM) and other ion-scale instabilities involving kinetic electrons \cite{Rosenbluth_PF_1968, Helander_PoP_2013, Proll_JPP_2022, Plunk_JPP_2017}. This property implies that the second adiabatic invariant $J$ is constant on magnetic surfaces and monotonically decreasing along the radius.
In figure \ref{fig:maximumJ}, depicting $J$ as a function of $\alpha$ and $r$ for CIEMAT-QI4 and for the standard configuration of W7-X, it can be observed that the maximum-$J$ property holds very closely in CIEMAT-QI4. The reason is that so-called flat mirror configurations (a notion introduced and developed in \cite{Velasco_NF_2023}) such as CIEMAT-QI4 tend to make $\partial_r J<0$ for all trapped particles even for low $\beta$ values. 

\begin{figure*}
	\begin{center}
		\includegraphics{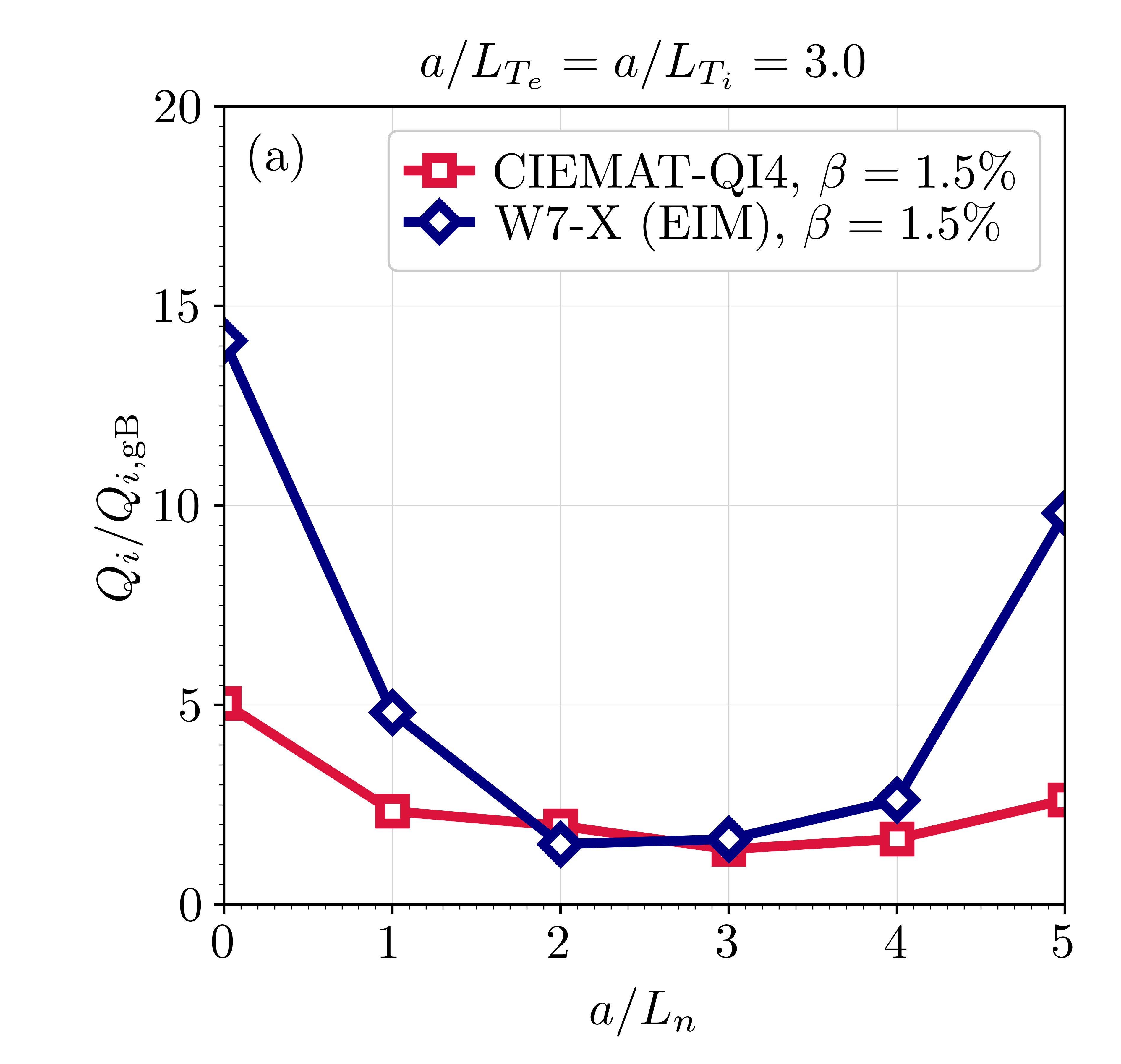}
		\includegraphics{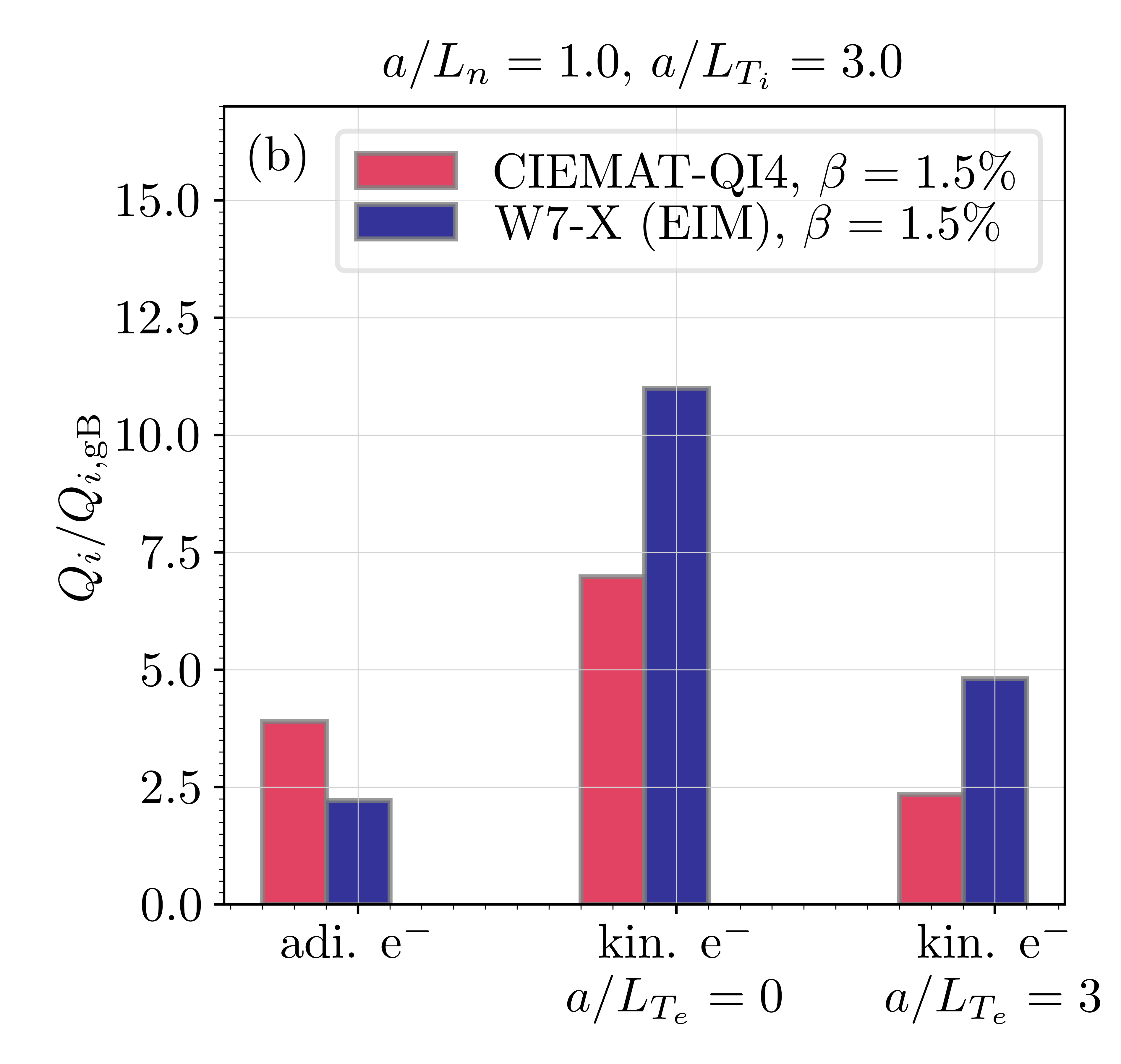}
	\end{center}
	\vspace{-4mm}
	\caption{(a) Ion heat flux as a function of the normalized density gradient for CIEMAT-QI4 
		(squares) and W7-X (diamonds), both configurations at $\beta=1.5\%$. (b) Ion heat flux for CIEMAT-QI4 and W7-X 
		in the case where $a/L_{n}=1.0$ and electrons are assumed adiabatic (left pair of bars), kinetic with vanishing temperature gradient (central pair of bars) and kinetic with finite temperature gradient (pair of bars on the right). Throughout this work the ion and electron heat fluxes are normalised to the gyro-Bohm heat flux of the ions $Q_{i,\mathrm{gB}}=(\rho_i/a)^2 n_i T_i v_{\mathrm{th},i}$}
	\label{fig:Qi_vs_fprim_and_bars}
\end{figure*}

In summary, the CIEMAT-QI4 geometric properties 
support the hypothesis of less detrimental turbulent transport and enhanced resilience to 
certain microinstabilities. This will be confirmed in section \ref{sec:turbulence}
by means of nonlinear gyrokinetic simulations.

\section{Turbulent transport in CIEMAT-QI4}
\label{sec:turbulence}

\begin{figure}
	\begin{center}
		\includegraphics{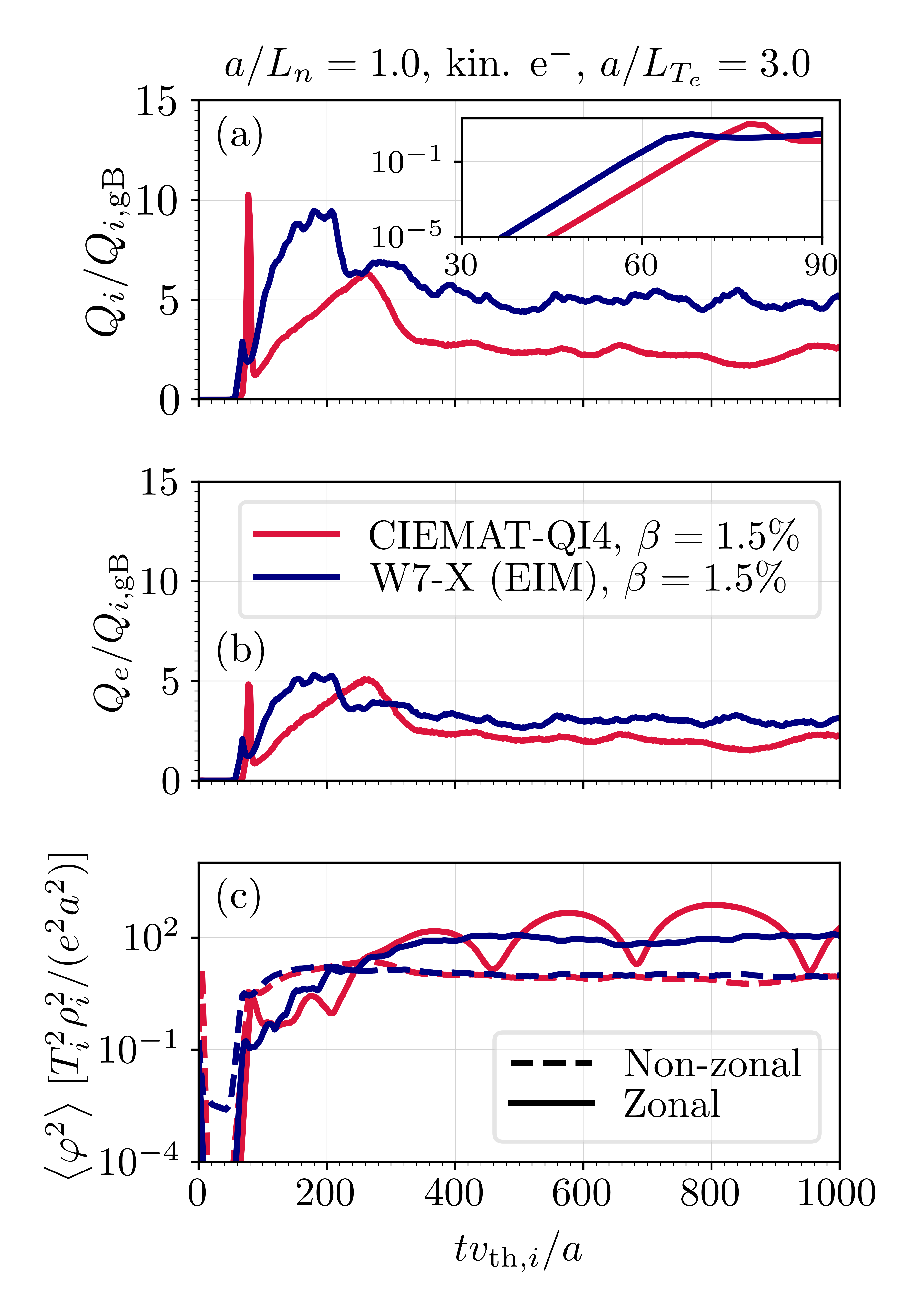}
	\end{center}
	\vspace{-4mm}
	\caption{\reftwo{For the case depiced in fig.~\ref{fig:Qi_vs_fprim_and_bars}(b) with $a/L_{n}=1$, $a/L_{T_i}=3$ and $a/L_{T_e}=3$, (a) shows the normalized ion heat flux as a function of time for CIEMAT-QI4 and W7-X. The inset shows the evolution of $Q_i$ in the linear phase on a logarithmic scale. For the same two configurations, (b) compares the normalized electron heat flux as a function of time, and (c) does so for the zonal and non-zonal components of the turbulent squared potential, averaged on the flux tube volume.}}
	\label{fig:saturation}
\end{figure}

Turbulence studies in stellarators have proliferated in recent years with the arrival of new codes
\cite{Barnes_jcp_391_2019,Maurer_JCP_2020,Mandell_arXiv_2022}, the increase in computational capability,
and the evidence that once neoclassical transport is optimised, turbulence comes to explain a good part
of experimental observations. For instance, taking \mbox{W7-X} as a paradigmatic case of the latter, turbulence simulations have explained 
why neither core density depletion \cite{Thienpondt_PRR_2023} nor the accumulation of impurities \cite{GarciaRegana_JPP_2021,GarciaRegana_NF_2021}, predicted by
neoclassical theory, take place in standard scenarios. Moreover, the remarkable reduction
of heat fluxes with increasing density gradient in \mbox{W7-X} \cite{Thienpondt_in_progress_2024}
aligns with the experimental evidence that increasing the bulk density peaking
is crucial to achieve high performance \cite{Bozhenkov_nf_2020}.
This section addresses the dependence of turbulent fluxes on the density gradient, by means of nonlinear \texttt{stella} simulations,  aiming to assess whether in CIEMAT-QI4 this advantageous characteristic of W7-X is preserved and even improved.

The nonlinear simulations performed with the code \texttt{stella}, are electrostatic, 
collisionless and consider a flux tube centered around the magnetic field line $ \alpha_0=0$, 
which lies on the flux surface at $r_0 /a = 0.7$ (the subindex 0 denotes quantities evaluated on the considered field line).
All simulations are performed with kinetic ions and electrons, unless stated otherwise.
For this radial position and field line,
the magnetic field strength and geometric coefficient $\mathcal{K}_\alpha$ for both configurations 
have already been presented in figure \ref{fig:configurations} and discussed in section \ref{sec:configurations}.

Both configurations have low global magnetic shear, $\hat{s}$, over practically their entire radius and, 
for the specific radial position chosen, $\hat{s}=-0.110$ for \mbox{W7-X} and $\hat{s}=-0.022$
for CIEMAT-QI4. Due to the low shear of both configurations, the
generalised twist and shift boundary conditions, proposed in \cite{Martin_ppcf_2018}, have been employed, extending
the flux tube up to $10.4$ and $9.0$ toroidal field periods in W7-X and CIEMAT-QI4, respectively, 
which corresponds to approximately two poloidal turns.
These lengths ensure that, considering the aforementioned boundary conditions, the width 
of the flux tube is equal along the radial, $x=(r-r_0)$, and binormal, $y=r_0(\alpha-\alpha_0)$, coordinates.
In other words, \refone{the smallest values of the perpendicular wavevector components} in these two directions are set to $k_{x,\mathrm{min}}=k_{y,\mathrm{min}}=0.067\rho_i^{-1}$ (which correspond to the lengths $L_x=L_y=94.2\rho_i$).
The smallest scales simulated are those with $k_{x,\mathrm{max}}=k_{y,\mathrm{max}}=3.0\rho_i^{-1}$.
With regard to the rest of the parameters of the simulations, \remark{the magnetic field line is sampled by $64$ points
per poloidal turn}, and the number of grid points in velocity space have been set to $N_{\mu}=12$
grid points in the magnetic moment, $\mu$, and $N_{v_{\|}}=64$ in the parallel velocity coordinate, $v_{\|}$.  
Here, $\rho_i=v_{\mathrm{th},i}/\Omega_i$ is the thermal ion Larmor radius, $v_{\mathrm{th},i}=\sqrt{2T_i/m_i}$ is the thermal speed, $\Omega_i=e B_r/m_i$ is the gyrofrequency, $e$ is the unit charge, $m_i$ is the ion mass, $T_i$ is the ion temperature and $B_r$ is a reference magnetic field (see \cite{Barnes_jcp_391_2019, GonzalezJerez_jpp_2022}, for further details on the electrostatic collisionless system of gyrokinetic equations solved by \texttt{stella}, coordinates used, geometry treatment and normalization conventions).

\begin{figure*}
	\begin{center}
		\includegraphics{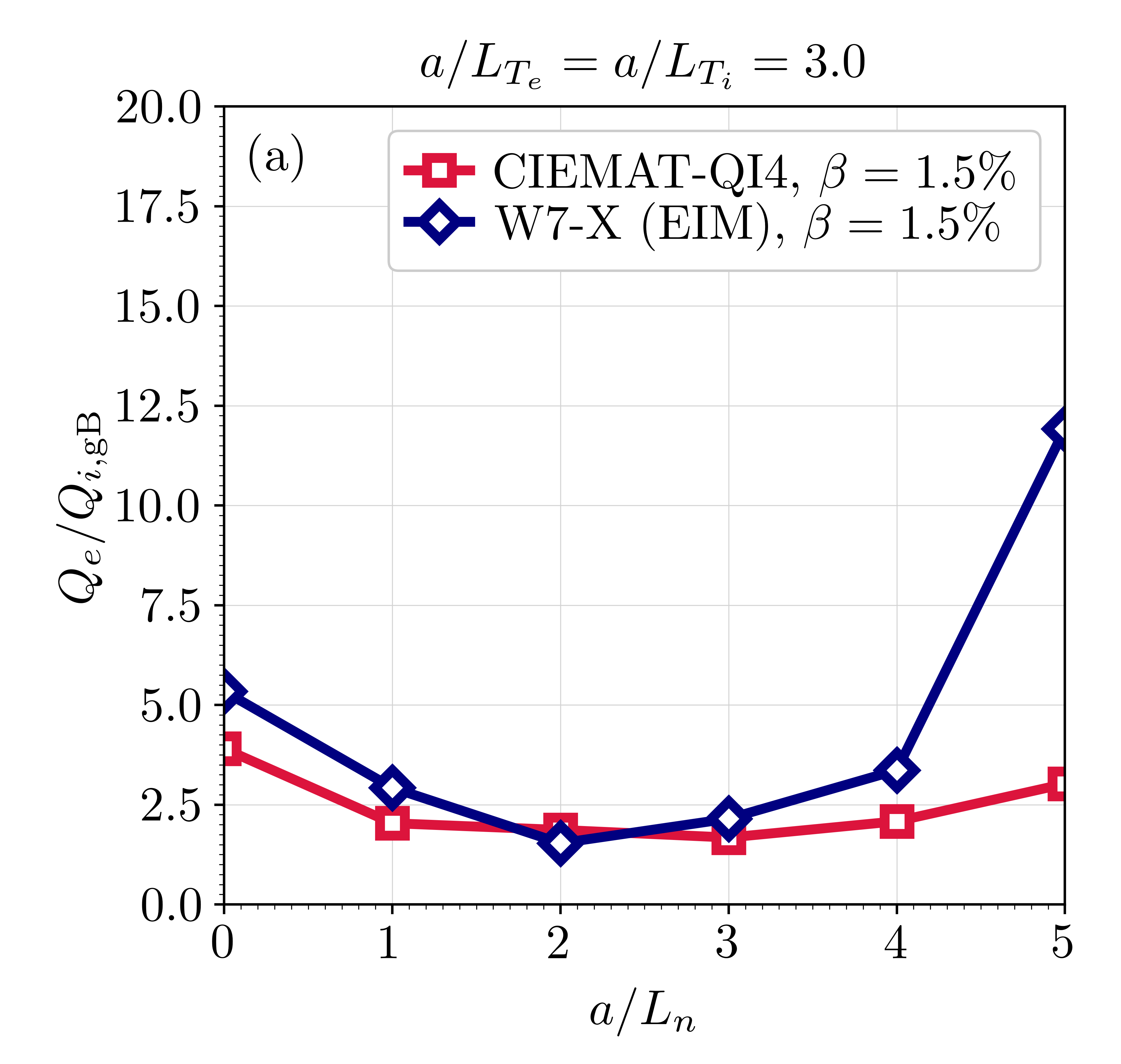}	
		\includegraphics{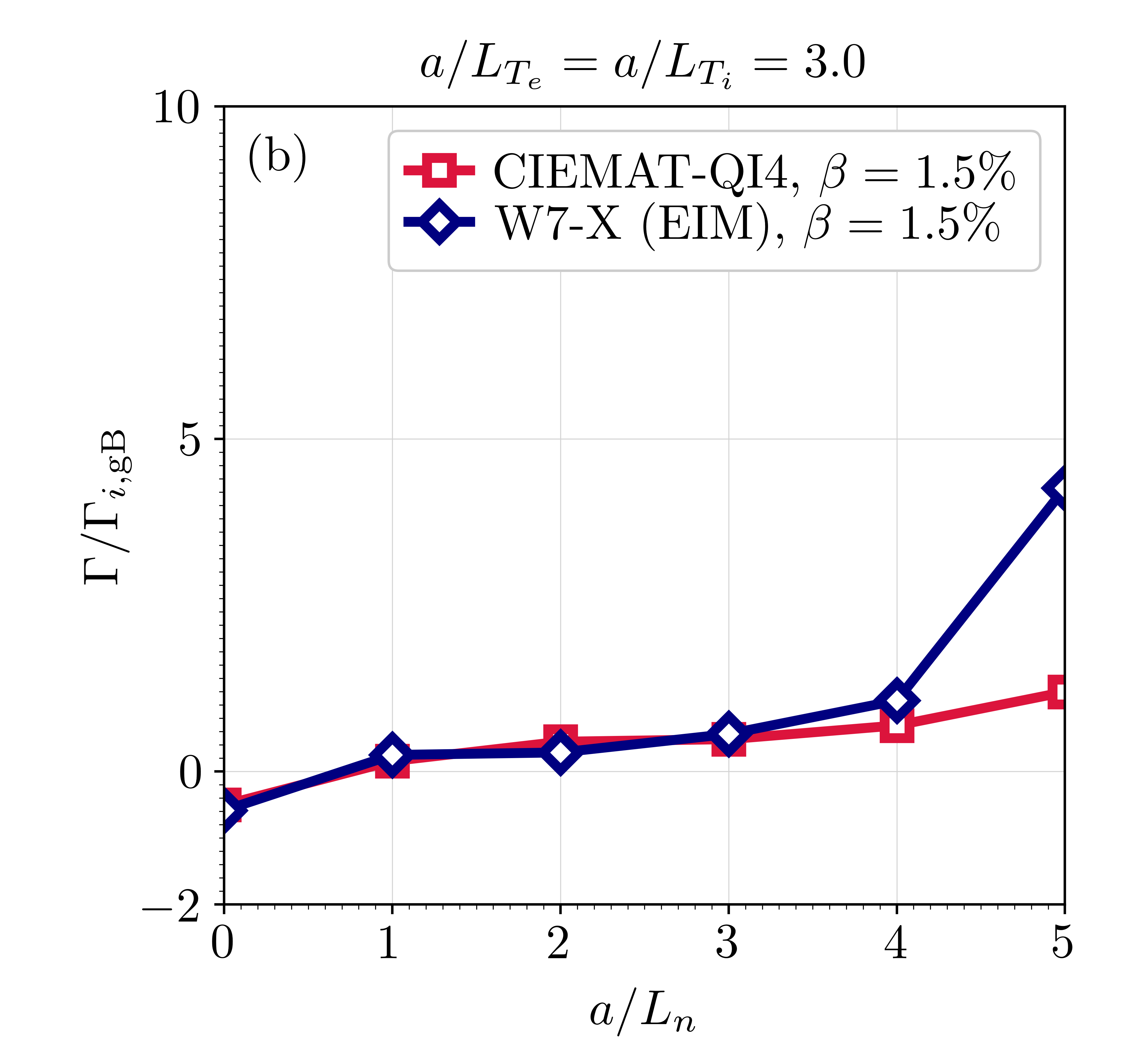}
	\end{center}
	\vspace{-4mm}
	\caption{(a) Electron heat flux and (b) particle flux as a function of the normalised density gradient for CIEMAT-QI4 (squares) and W7-X standard configuration (diamonds), both at $\beta=1.5\%$. The particle flux is normalised to the gyro-Bohm value of the ions, $\Gamma_{i,\mathrm{gB}}=n_i v_{\mathrm{th},i}(\rho_i/a)^2$.}
	\label{fig:Qe_Gamma_vs_fprim}
\end{figure*}

\begin{figure*}
	\centering
	\includegraphics{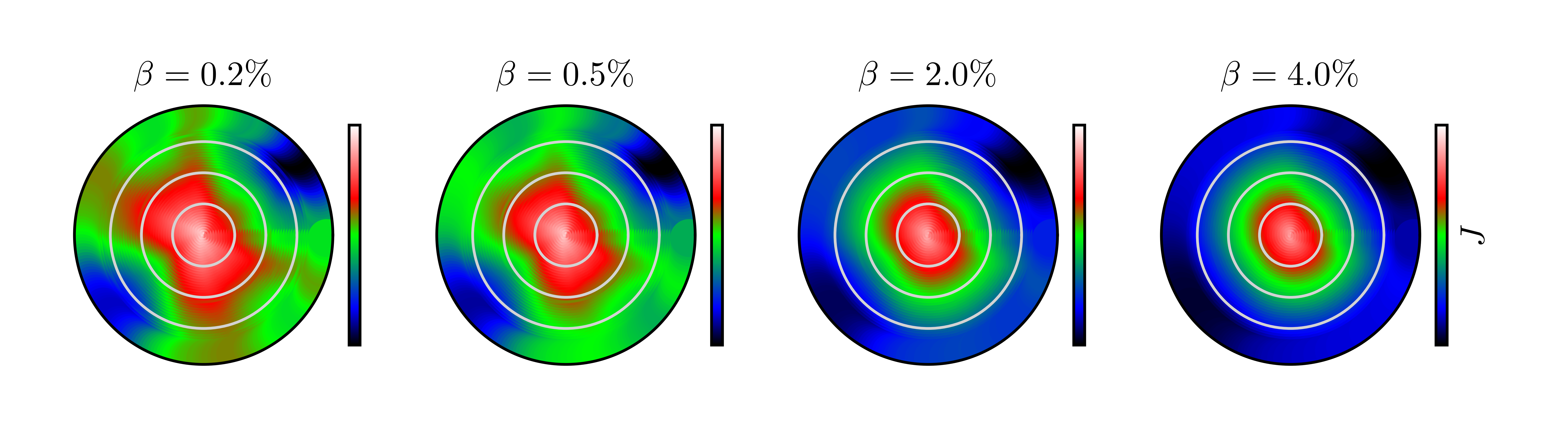}
	\vspace{-4mm}
	\caption{Second adiabatic invariant, $J$, for CIEMAT-QI4 at different values of $\beta $. 
		For this representation, $(r/a)^2$ and $\alpha$ are used as radial and angular, respectively, polar coordinates. The circular
		contours represent the flux surfaces $(r/a)^2=\{0.25, 0.50, 0.75\}$.}
	\label{fig:maps_of_J_beta_scan}
\end{figure*}

As we said above, we have carried out a scan in the density gradient,
$a/L_{n}$, at finite electron and ion temperature gradients $a/L_{T_i}=a/L_{T_e}=3.0$.
Here, $a/L_{X}=-(a/X)\mathrm{d}X/\mathrm{d}r$ is the characteristic radial variation length scale, 
also referred to as the normalised gradient,
of a given plasma profile $X(r)$. 
At the selected radial position, temperature gradients are typically around the values previously indicated in standard ECRH and ECRH+NBI W7-X plasmas, whereas in enhanced performance discharges they can be larger. Regarding $a/L_n$, W7-X exhibits core values in the range of approximately $0.5-3$, depending on the scenario \cite{Carralero_nf_2021}. However, recent discharges have shown larger values during pure NBI phases \cite{Romba_nf_2023}.

The results of these scans in $a/L_n$ for the ion heat flux, $Q_i$,
are shown in figure \ref{fig:Qi_vs_fprim_and_bars}(a) for the two configurations.
In this figure, one can clearly observe how CIEMAT-QI4 at low density gradients
exhibits appreciably lower $Q_i$ values than in W7-X. 
Specifically, for the case $a/L_{n}=0$, the ion heat flux is approximately $2.8$ 
times lower than for W7-X, and for $a/L_{n}=1$, the reduction reaches a factor of nearly $2.0$.
Towards higher density gradient values ($a/L_n=0\rightarrow2$), W7-X exhibits the $Q_i$ reduction
reported in \cite{Thienpondt_in_progress_2024, Thienpondt_ISHW_2022}, 
following a mild increase of $Q_i$ in the transition $a/L_n=2\rightarrow 4$, 
and a very pronounced growth for $a/L_n=4\rightarrow5$. 
In contrast, CIEMAT-QI4 maintains a consistently low ion heat flux throughout the scanned density gradients. 
Thus, the ion heat flux level is low and robust to increasing $a/L_n$, with only a mild increase
when $a/L_n$ becomes the largest in all the represented range. 
In summary, CIEMAT-QI4 starts with significantly lower $Q_i$ than W7-X when the density gradient is low, is comparable to \mbox{W7-X} at intermediate $a/L_n$ values, and withstands the impact of moderate to very high values of $a/L_n$.

\begin{figure*}
	\centering
	\includegraphics{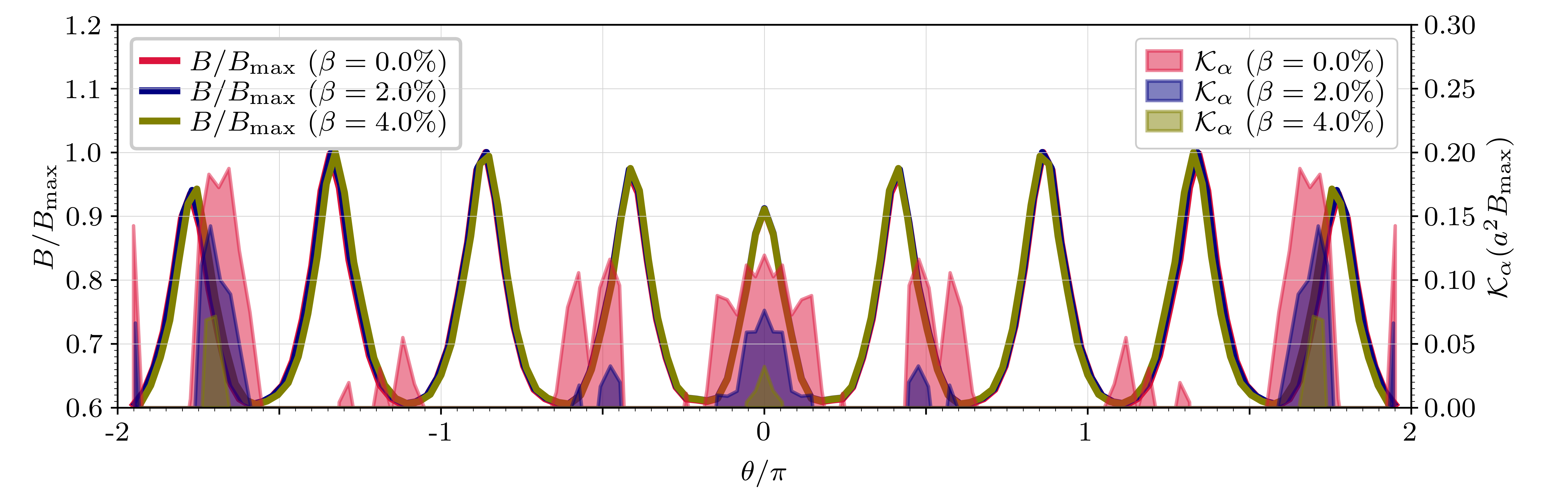}
	\vspace{-4mm}
	\caption{Regions of bad curvature ($\mathcal{K}_{\alpha}>0$) along the poloidal coordinate of the simulated flux tube for CIEMAT-QI4 at $\beta=0\%$ (red areas), $2\%$ (blue areas), and $4\%$ (green areas).}
	\label{fig:bad_curvature_beta}
\end{figure*}

\begin{figure*}
	\begin{center}
		\includegraphics{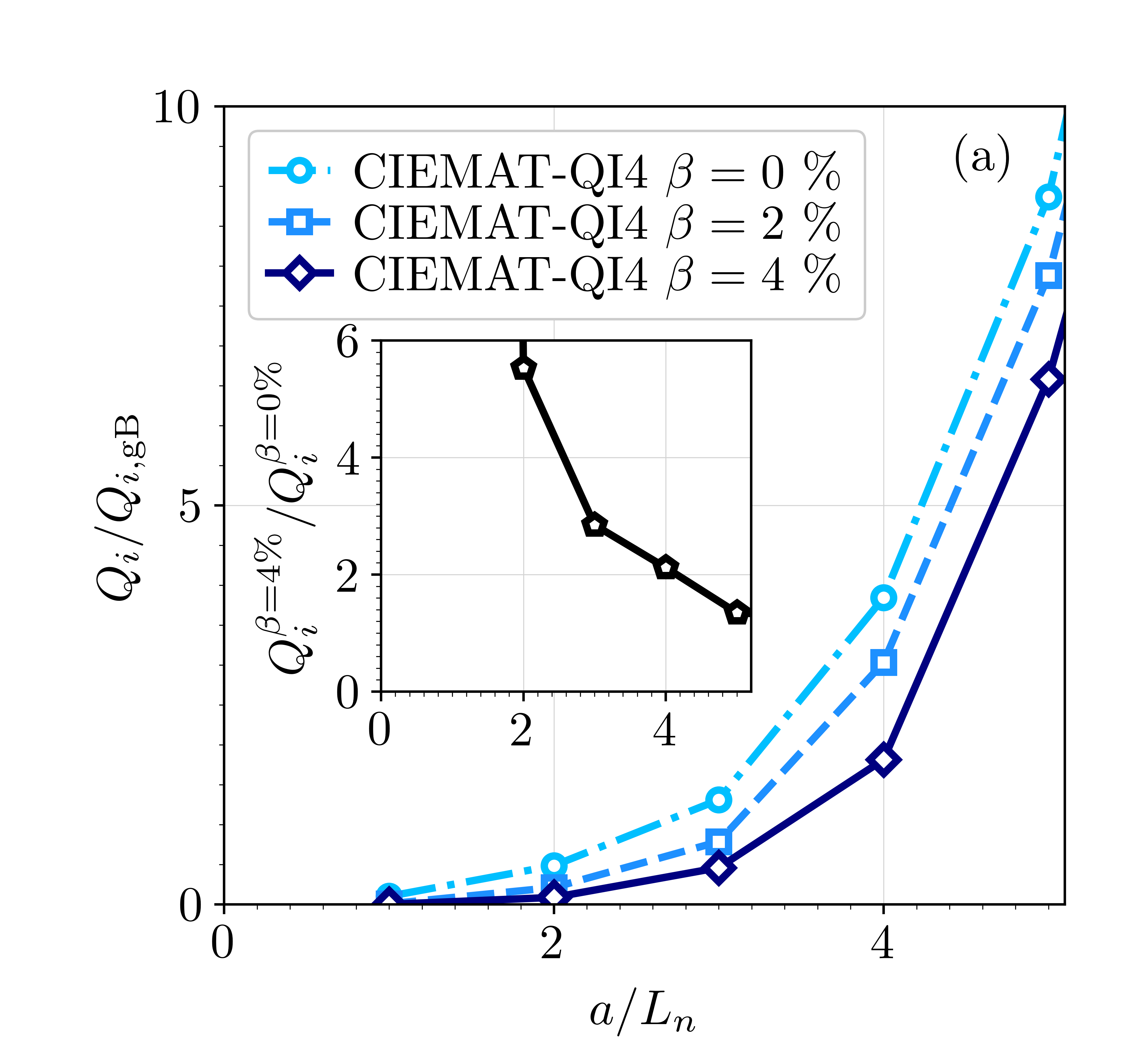}
		\includegraphics{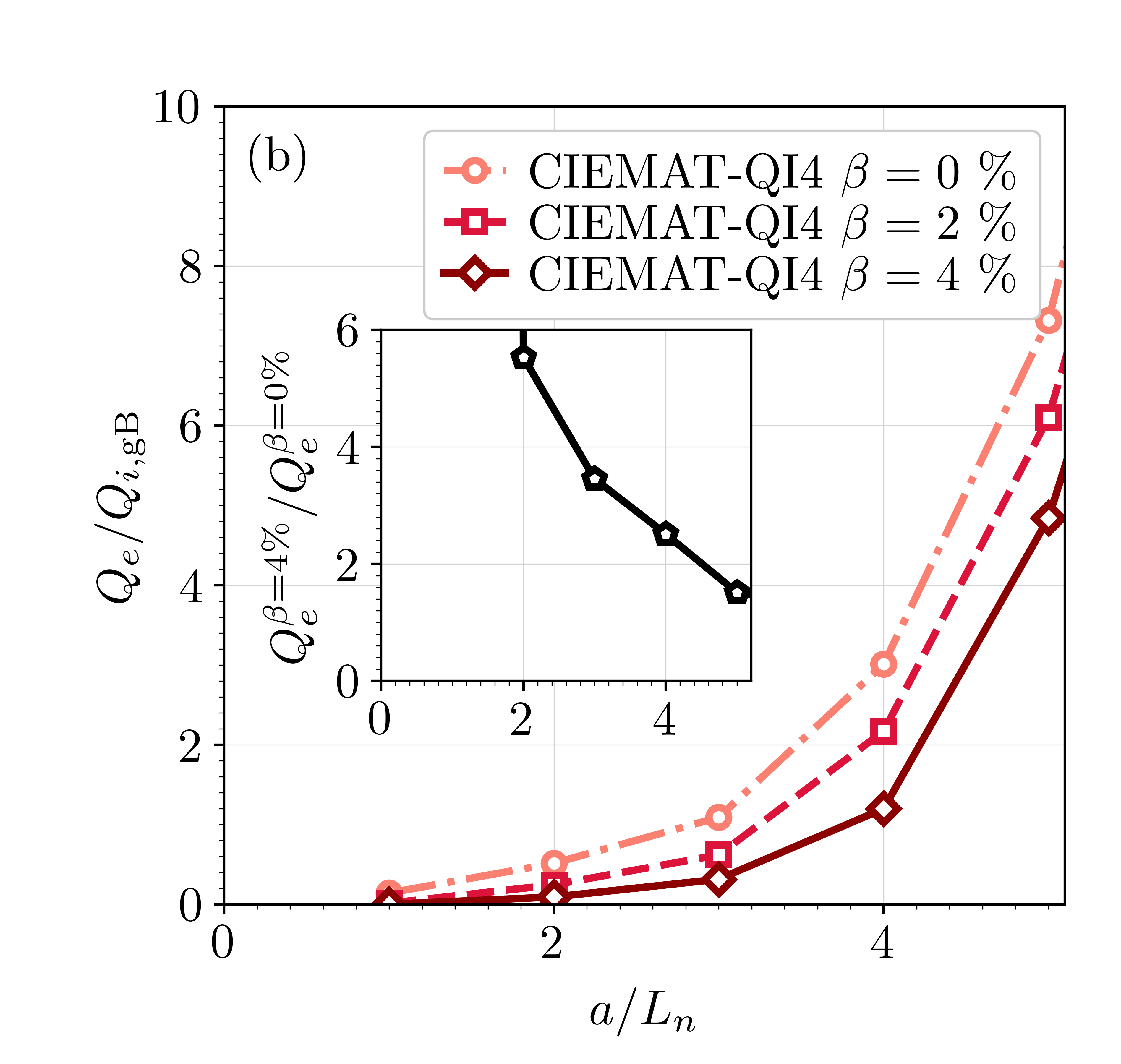}
	\end{center}
	\vspace{-4mm}
	\caption{Main ion (a) and electron (b) heat flux as a function of the normalised density gradient for vanishing temperature gradient for both species, considering the CIEMAT-QI4 configuration at $\beta=0\%$ (circles), $2\%$ (squares), and $4\%$ (diamonds). The inset in each figure depicts the ratio of their values at $\beta=4\%$ to those at $\beta=0\%$ for the corresponding ion or electron heat flux.}
	\label{fig:Qs_vs_fprim_beta_scan}
\end{figure*}

The better performance of CIEMAT-QI4 against W7-X for moderate to high density gradients, away from the low density gradient region where ITG is most unstable, is likely due to the diminsihed role predicted by theory of electron-driven instabilities in configurations that approach the maximum-$J$ property.
However, it is important to note that in the low density gradient region, the comparison just discussed involves electrons as well. Focusing on the case $a/L_n=1$, in figure \ref{fig:Qi_vs_fprim_and_bars}(b) the ion heat flux is shown in the case where electrons are considered adiabatic, kinetic with zero temperature gradient ($a/L_{T_e}=0$), and kinetic with finite temperature gradient ($a/L_{T_i}=a/L_{T_e}=3.0$). It can be seen that both W7-X and CIEMAT-QI4 experience an increase in ion heat flux when kinetic electrons are added, even with vanishing electron temperature gradient ($a/L_{T_e}=0$). Among the two devices, the increase in the ion heat flux, already known in the literature \cite{Proll_JPP_2022}, is much more pronounced in W7-X. Adding the temperature gradient to kinetic electrons makes the heat flux decrease, much more for CIEMAT-QI than for W7-X. \\

\reftwo{The role of a finite electron temperature gradient has been recently investigated for the accurate quantification of turbulent fluxes in W7-X \cite{Thienpondt_PRR_2023, Thienpondt_in_progress_2024, Zocco2024, Wilms2024}. In tokamaks, $a/L_{T_e}$ is related ot the excitation of temperature gradient driven TEMs at ion Larmor scales (see e.g.~\cite{Hillesheim2013, Li2022}) which may play a role in the specific case under discussion and cannot be entirely ruled out.}
\reftwo{Looking at the time traces of our simulations for the specific case with finite $a/L_{T_e}$ and $a/L_n=1$, 
	corresponding to the rightmost bars of figure \ref{fig:Qi_vs_fprim_and_bars}(b), we can see a saturation of the turbulence with distinctive characteristics in both devices. Figure \ref{fig:saturation}(a) represents the ion heat flux as a function of time. Firstly, no appreciable difference is observed in the linear phase, represented in the inset. The linear growth rates appear very similar in CIEMAT-QI4 and W7-X. However, after the linear phase, a slower growth of $Q_i$ makes it reach a lower level in CIEMAT-QI4 than in W7-X (the evolution of $Q_e$, represented in the figure \ref{fig:saturation} (b), follows an evolution similar to that of $Q_i$). After that, a decrease in heat fluxes is observed at $t v_{\mathrm{th},i}/a\approx 200$ for W7-X and, more noticeably, at $t v_{\mathrm{th},i}/a\approx 300$ for CIEMAT-QI4. In both devices, that decrease correlates with an increase in the zonal component of the potential, see \ref{fig:saturation}(c), which, on the other hand, shows clear differences along	 their time evolution. While the zonal potential in W7-X reaches a saturated value that barely varies, large amplitude and long period oscillations are observed in CIEMAT-QI4, eventually exceeding its value in W7-X.}

Turning back to the density gradient scan with kinetic electrons and all drives, in figures \ref{fig:Qe_Gamma_vs_fprim}(a) and \ref{fig:Qe_Gamma_vs_fprim}(b), the electron heat ($Q_e$) and particle ($\Gamma$) fluxes, respectively, are presented. 
Both figures show a difference at low density gradient values not as significant as for the ion heat flux (figure \ref{fig:Qi_vs_fprim_and_bars}(a)) but, still visible for $a/L_n\le 1$ and $Q_e$. Again, as the density increases above $a/L_n=3$ the curves of both devices separate from each other, with CIEMAT-QI4 maintaining lower levels of $Q_e$ and $\Gamma$. As for $\Gamma$ both configurations find a comparable level of inward turbulent flux at zero density gradient (turbulent pinch), that has been proven to counteract effectively the neoclassical thermo-diffusion responsible for the tendency to density core depletion in stellarators \cite{Thienpondt_PRR_2023}. As the density gradient increases, the particle flux increases at a comparable rate up to $a/L_n=3$, after which CIEMAT-QI4 maintains a substantially lower particle flux than W7-X. The latter is important in view of the capacity of a configuration to support the formation of a density pedestal, an ingredient with a recognised role in the generation of transport barriers and access to high confinement mode.

\begin{figure}
	\begin{center}
		\includegraphics{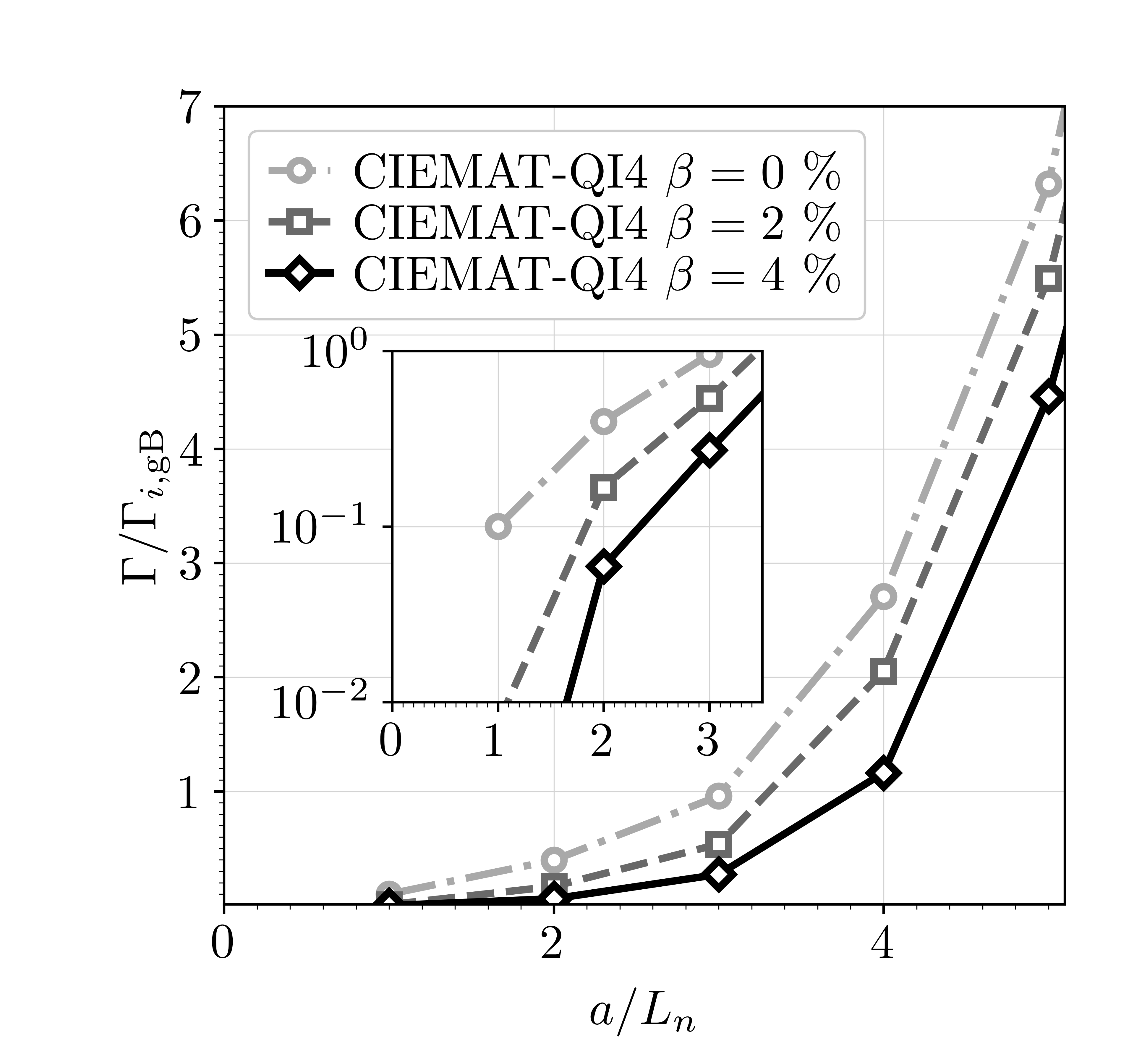}
	\end{center}
	\vspace{-4mm}
	\caption{Turbulent particle flux as a function of the normalised density gradient for vanishing ion and electron temperature gradients, considering CIEMAT-QI4 at $\beta=0\%$ (circles), $2\%$ (squares), and $4\%$ (diamonds). The inset displays a logarithmic scale zoom focusing on the lowest values of the density gradient within the scanned range.}
	\label{fig:Gamma_vs_fprim_beta_scan}
\end{figure}

Delving into the correlation between the proximity to maximum-$J$ and
turbulent transport, a scan has been performed on the density gradient, 
considering CIEMAT-QI4 at three different values of the normalised plasma pressure, $\beta$. 
Prior to the presentation of the results, it is important to recall that 
the configuration gets closer to exactly satisfying the maximum-$J$ property as $\beta$ increases. 
This can be observed in figure \ref{fig:maps_of_J_beta_scan}.
We will focus on turbulent transport driven exclusively by the density gradient because theory predicts that density-gradient-driven TEM are stable in exactly maximum-$J$ configurations \cite{Rosenbluth_PF_1968, Helander_PoP_2013}. Thus, vanishing values of the ion and electron 
temperature gradients have been taken. Looking at the coefficient $\mathcal{K}_{\alpha}$ along the parallel coordinate 
of the simulated flux tubes, represented in the
the figure \ref{fig:bad_curvature_beta}, one can also observe
that the differences between CIEMAT-QI4 at different values of $\beta$ are very pronounced 
for this parameter in particular. 
Not only for the transition from $\beta=0\%$ to $2\%$ the areas of bad curvature shrink, but also 
for the transition from $\beta=2\%$ to $4\%$ \refone{(it is important to note that, aside from the correlation between 
the maximum-$J$ degree and $\beta$ explored in this part of the study, 
for the case of $\beta=4\%$, the electrostatic limit assumed in this work may be questionable)}.
The results of this scan are shown in figures \ref{fig:Qs_vs_fprim_beta_scan}(a)
for the ion heat flux, figure \ref{fig:Qs_vs_fprim_beta_scan}(b) for the electron heat flux and 
figure \ref{fig:Gamma_vs_fprim_beta_scan} for the particle flux. \reftwo{We note for the reader that by setting the ion and electron temperature gradients to zero, we isolate purely convective sources of ion and electron heat transport as well as the diffusive component of particle transport}.

The three fluxes follow apparently an exponential increase with the density gradient\reftwo{,
being rather low for $a/L_{n}\lesssim 2$. This suggests that the heat fluxes observed at finite temperature gradients in figs.~\ref{fig:Qi_vs_fprim_and_bars}(a) and \ref{fig:Qe_Gamma_vs_fprim}(a)  for $a/L_{n}\lesssim 2$ are primarily diffusive, assuming that the ion heat flux produced by the electron temperature gradient and vice versa are negligible in that region of the parameter space}.
It is also observed that, indeed, the increase in $\beta$ produces significantly
lower fluxes. Looking at the
insets of figures \ref{fig:Qs_vs_fprim_beta_scan}(a) and \ref{fig:Qs_vs_fprim_beta_scan}(b),
which represent the ratio between the corresponding heat flux at $\beta=0\%$ and $\beta=4\%$,
it is observed that the difference factor between the two becomes increasingly larger
with decreasing $a/L_n$. For $a/L_n=5$, the heat fluxes at $\beta=4\%$ are $1.5$ 
lower than for $\beta=0\%$, while that factor increases up to nearly $6$ for $a/L_n=2$.
Because the case $\beta=4\%$ is marginally unstable at $a/L_n=1$ the ratio
$Q^{\beta=0\%}_s/Q_s^{\beta=4\%}$ approaches off-scale values near to one hundred.
Finally, looking at the particle flux, $\Gamma$, the exponential trend with $a/L_n$,
common to those just discussed for $Q_i$ and $Q_e$, points out the stiff character of particle transport.
Analogously to the scaling derived for the heat flux versus the temperature gradient
from critical balance arguments \cite{Barnes_prl_2011},
the trend of the particle flux versus the density gradient can be quantified.
Fitting the particle flux to the functional form $\Gamma=K_1(a/L_n)^{K_2}$ 
results in exponents of around $K_2\sim 4$, see table \ref{tab:Gamma_stiffness}
for the specific values of the coefficients $K_1$ and $K_2$ for each case of $\beta$.
For this fitting, only the cases with $a/L_n\ge 2$ of well developed 
turbulence have been considered.
The particle flux at lower gradients, isolated in 
figure \ref{fig:Gamma_vs_fprim_beta_scan} (inset) using logarithmic scale, shows how 
$\Gamma(a/L_n)$ tends to significantly critical gradients larger as $\beta$ increases. 
In summary, the simulations for CIEMAT-QI4 at different values of $\beta$ show that the greater the value of $\beta$ and,
consequently, the closer to maximum-$J$, the lower the transport driven by electrostic turbulence and the larger its critical density gradient.




\begin{table}[h]
	\centering
	\begin{tabular}{c c c}
		\hline
		$\beta$ [$\%$] & $K_1$ & $K_2$ \\
		\hline
		0.0 & $(1.3\pm 0.2)\times 10^{-2}$ & $3.85\pm 0.09$ \\
		2.0 & $(0.7\pm 0.1)\times 10^{-2}$ & $4.12\pm 0.09$ \\
		4.0 & $(0.2\pm 0.1)\times 10^{-2}$ & $4.7\pm 0.3$ \\
		\hline
	\end{tabular}
	\caption{Results from fitting the particle fluxes with \mbox{$a/L_n\le 3$} represented in figure \ref{fig:Gamma_vs_fprim_beta_scan} to a function with the form $\Gamma=K_1(a/L_{n})^{K_2}$.}
	\label{tab:Gamma_stiffness}
\end{table}

\section{Conclusions}
\label{sec:conclusions}
CIEMAT-QI4 is a quasi-isodynamic configuration that approximately satisfies the maximum-$J$ condition even at small plasma $\beta$. CIEMAT-QI4 was introduced in \cite{Sanchez_NF_2023}, where it was shown that it exhibits very good fast-ion confinement for a broad range of $\beta$ values, gives reduced neoclassical transport and small bootstrap current,  and is ideal MHD stable up to $\beta=5 \%$. In the present paper we have reported a first numerical analysis of turbulent transport for the CIEMAT-QI4 configuration by means of nonlinear electrostatic gyrokinetic simulations performed with the code \texttt{stella}~\cite{Barnes_jcp_391_2019}.

A scan of the turbulent heat fluxes in the density gradient, keeping finite values of the ion and electron temperature gradients, shows that CIEMAT-QI4 features: 1) low turbulent heat fluxes at flat or weakly peaked density profiles. In particular, it has been shown that the configuration is resilient to the strong increase in the heat flux typically produced by introducing  kinetic electrons~\cite{Thienpondt_in_progress_2024, Proll_JPP_2022};~2) the turbulent heat fluxes remain low and weakly dependent on the density gradient at moderate values thereof; 3) the fluxes increase very mildly when going to strongly peaked density profiles, as expected from analytical arguments relating resistance to density-gradient-driven trapped-electron-mode instabilities and to the proximity to the maximum-$J$ property \cite{Rosenbluth_PF_1968, Helander_PoP_2013}. In summary, CIEMAT-QI4 exhibits reduced turbulent transport in wide experimentally relevant regions of parameter space.

In addition, turbulent transport driven solely by density-gradient-driven trapped-electron-modes has been investigated in CIEMAT-QI4 for different values of $\beta$, (motivated, on the one hand, by the fact that in quasi-isodynamic configurations, and therefore in CIEMAT-QI4, the larger $\beta$ the closer the configuration is to satisfying the maximum-$J$ property. And, on the other hand, by the analytical results \cite{Rosenbluth_PF_1968, Helander_PoP_2013} that predict stability of density-gradient-driven trapped-electron-modes for exactly maximum-$J$ configurations). Although with comparable stiffness, turbulent fluxes are significantly lower at higher $\beta$ and tend to larger critical gradients.

In the future, \remark{it will be necessary to study how the reduction in the turbulent fluxes reported in this paper} translates into the equilibrium profiles determined by transport calculations \cite{BanonNavarro_NF_2023} and \remark{how electromagnetic effects modify the results at high $\beta$ \cite{Aleynikova_jpp_2018, Wilms_jpp_2021}}.

\section*{Acknowledgements}
\label{sec:acknowledgements}
J.M.G.R. is grateful to A. Bañón Navarro for fruitful discussions and Michael Barnes for his support with the code \texttt{stella}. This work has been carried out within the framework of the EUROfusion Consortium, funded by the European Union via the Euratom Research and Training Programme (Grant Agreement No 101052200 – EUROfusion). Views and opinions expressed are however those of the author(s) only and do not necessarily reflect those of the European Union or the European Commission. Neither the European Union nor the European Commission can be held responsible for them. This research was supported in part by grant PID2021-123175NB-I00, Ministerio de Ciencia e Innovación, Spain. Simulations were performed in the supercomputers Marconi (CINECA, Italy) under the \texttt{FUA\_STELTURB} project of the \textit{TSVV\#13 Task Stellarator Turbulence Simulation}.

\section*{Bibliography}
\bibliographystyle{unsrt}
\bibliography{./biblio}

\end{document}